\documentstyle[12pt]{article}
\def\CA{{\cal A}}    
 \def\CF{{\cal F}}  \def\CH{{\cal H}} 
\def\CI{{\cal I}}    
 \def\CN{{\cal N}} \def\CO{{\cal O}}

\def\BA{{\bf A}}   \def\BD{{\bf D}} 
 \def\BF{{\bf F}} \def\BG{{\bf G}}  
 \def\BJ{{\bf J}} \def\BK{{\bf K}} \def\BL{{\bf L}} 
 \def\BN{{\bf N}}  
  \def\BS{{\bf S}} 
   \def\BX{{\bf X}}
\def\BY{{\bf Y}} 
\def\Ba{{\bf a}} \def\Bb{{\bf b}}

  \def\Bo{{\bf o}} 
   
\def\Bu{{\bf u}}   \def\Bx{{\bf x}}

\def\proof{\medskip\noindent{\sl Proof\ } :\  }

\def\condition#1
{\medskip\noindent{\bf Condition\ } #1 :\ }

\def\proposition#1{\medskip\noindent{\bf Proposition\ } #1 :\ }

\def\qed{\hfill{\sl q.e.d.} \medskip}

\def\and{\quad{\rm and}\quad}

\def\where{\quad{\rm where}\quad}

\def\half{{1\over2}}
\def\p{\partial}
\def\Tr#1{{\rm T}\!{\rm r}\{#1\}}

\def\tr{{\rm T}\!{\rm r}}

\def\id{{\bf1}}
\def\mat#1{\left(\matrix{#1}\right)}

\def\IR{\relax{\rm I\kern-.18em R}}

\def\pr0#1#2#3{{\it Phys.\ Rev.} {\bf #1} (19#2) #3}

\def\boxit#1{\hfil\break\vbox{\hrule\hbox{\vrule\kern3pt\vbox{\kern3pt#1\kern3pt}\kern3pt\vrule}\hrule}}

\def\complex{{\mathchoice
{\setbox0=\hbox{$\displaystyle\rm C$}\hbox{\hbox to0pt
{\kern0.4\wd0\vrule height0.9\ht0\hss}\box0}}
{\setbox0=\hbox{$\textstyle\rm C$}\hbox{\hbox to0pt
{\kern0.4\wd0\vrule height0.9\ht0\hss}\box0}}
{\setbox0=\hbox{$\scriptstyle\rm C$}\hbox{\hbox to0pt
{\kern0.4\wd0\vrule height0.9\ht0\hss}\box0}}
{\setbox0=\hbox{$\scriptscriptstyle\rm C$}\hbox{\hbox to0pt
{\kern0.4\wd0\vrule height0.9\ht0\hss}\box0}}}}
\def\Co{{\mathchoice
{\setbox0=\hbox{$\displaystyle\rm C$}\hbox{\hbox to0pt
{\kern0.4\wd0\vrule height0.9\ht0\hss}\box0}}
{\setbox0=\hbox{$\textstyle\rm C$}\hbox{\hbox to0pt
{\kern0.4\wd0\vrule height0.9\ht0\hss}\box0}}
{\setbox0=\hbox{$\scriptstyle\rm C$}\hbox{\hbox to0pt
{\kern0.4\wd0\vrule height0.9\ht0\hss}\box0}}
{\setbox0=\hbox{$\scriptscriptstyle\rm C$}\hbox{\hbox to0pt
{\kern0.4\wd0\vrule height0.9\ht0\hss}\box0}}}}
\def\Rl{{\mathchoice
{\setbox0=\hbox{$\displaystyle\rm R$}\hbox{\hbox to0pt
{\kern0.4\wd0\vrule height0.9\ht0\hss}\box0}}
{\setbox0=\hbox{$\textstyle\rm R$}\hbox{\hbox to0pt
{\kern0.4\wd0\vrule height0.9\ht0\hss}\box0}}
{\setbox0=\hbox{$\scriptstyle\rm R$}\hbox{\hbox to0pt
{\kern0.4\wd0\vrule height0.9\ht0\hss}\box0}}
{\setbox0=\hbox{$\scriptscriptstyle\rm R$}\hbox{\hbox to0pt
{\kern0.4\wd0\vrule height0.9\ht0\hss}\box0}}}}

\def\be{\begin{equation}}
\def\ee{\end{equation}}
\def\bea{\begin{eqnarray}}
\def\eea{\end{eqnarray}}
\def\mbegin#1{\left(\begin{array}{#1}}
\def\mend{\end{array}\right)}

\def\benu{\begin{enumerate}}
\def\eenu{\end{enumerate}}

\def\bbib{\begin{thebibliography}{99}}
\def\ebib{\end{thebibliography}}

\newfont{\frak}{eufm10 scaled 1200}

\newfont{\Bbb}{eusb10 scaled 1200}

\begin{document} 

\begin{titlepage}
\vspace*{-4ex}

%%%%%%%%%%%% write the preprint number and datum %%%%%%%%
\null \hfill Preprint TU-528  \\
\null \hfill January 1998 \\
\null \hfill revised: December 1999 \\[2ex]

%%%%%%%%%%%% here starts the title

\begin{center}
{\LARGE\bf Noncommutative Geometry \\[1em] 
and \\[1em] 
Gauge Theory on Fuzzy Sphere}\\[5em] 
%%%%%%%%%% Here is the title of this Note

%%%%%%%%%%%%%%%%%% here authors name %%%%%%%%%%%%
Ursula Carow-Watamura\footnote{
e-mail: ursula@tuhep.phys.tohoku.ac.jp} \  and \  
Satoshi Watamura\footnote{e-mail: watamura@tuhep.phys.tohoku.ac.jp}

%%%%%%%%%%%%%%%%%%%%%%%%%%%%%% address of the institute %%%%%%%%
Department of Physics \\
Graduate School of Science \\
Tohoku University \\
Aoba-ku, Sendai 980-8577, JAPAN \\ [2ex]
\end{center}

\bigskip

\begin{abstract}

The differential algebra on the fuzzy sphere 
is constructed by applying Connes' scheme.
The $U(1)$ gauge theory on the fuzzy sphere based on this 
differential algebra is defined.
The local $U(1)$ gauge transformation on the fuzzy sphere is
identified with the left $U(N+1)$ transformation of the field,
where a field is a bimodule over the
quantized algebra $\CA_N$.
The interaction with a complex scalar field is 
also given.

\end{abstract}

\end{titlepage}

\eject  %%%%% Title page is over

\section{Introduction}

The concept of quantized spaces is discussed in a variety of 
fields in physics and mathematics. 
From the physicists' viewpoint, the main motivation for investigating 
noncommutative spaces stems from the need of 
an appropriate framework to describe the quantum theory of 
gravity. Recently quantized spaces are also discussed 
in connection with M(atrix) theory 
which has been proposed as a nonperturbative formulation 
of string theory\cite{BFSS,IKKT}. 
This development in string theory supports the idea that
the noncommutative structure of spacetime becomes relevant 
when constructing the theory of gravitation 
at Planck scale. 

To describe noncommutative spaces, 
the noncommutative geometry is now investigated by many authors and 
using this framework one can even consider the differential geometry
of singular spaces like, for example, a $2$-point space 
which has been shown to provide a geometrical interpretation 
of the Higgs mechanism\cite{ConnLott90}. 

On the other hand, 
in order to describe gravity we have to know the theory of a 
wider class of noncommutative geometry.
In this context, the class of noncommutative spaces 
which can be considered as 
deformations of continuous spaces is especially interesting. 
In general, such noncommutative spaces can be obtained by quantizing 
a given 
 space with its Poisson structure.  Furthermore, if the original space is
compact one obtains a finite dimensional matrix algebra as a 
quantized algebra
of  functions over this space.
In this case, we may consider the deformation as 
a kind of regularization with the special property that 
we can keep track of the geometric structure, a feature which 
is missing in the conventional 
regularization schemes.

In physics the algebra of the fuzzy sphere is well known 
and has been investigated in a variety of 
contexts: as an example for a general quantization 
procedure \cite{Berezin74,Berezin75a} (see also for example 
\cite{BordHopp91,BordMein94,Coburn92,KlimLes92,RawnCahen90} 
 and references therein) and in relation with geometric quantization. 
It is also discussed as the algebra appearing in membranes 
\cite{Hoppe89,deWiHopp88}, 
in relation with coherent states \cite{Bargmann61,Perelomov72}, 
and recently 
in connection with noncommutative geometry 
\cite{MadoreCQG92,GrosMado95,GrosKlim95}. The same structure also 
appears 
in the context of the quantum Hall effect \cite{Haldane83,FanoOrto86}.
In this paper, we investigate the differential geometry of the fuzzy sphere and the field theory on it. We formulate the 
$U(1)$ gauge theory on the fuzzy sphere. 
The fuzzy sphere is one example in 
the above mentioned class of noncommutative geometry and
thus the field theory on this space is a very instructive model 
to examine the ideas of noncommutative geometry.
Besides that, it is a deformation of the sphere 
obtained by quantization based on the Poisson structure on $S^2$, 
and the resulting algebra $\CA_N$ is a finite dimensional matrix algebra. 
Thus, what we obtain is a regularized field theory on the sphere.
From this point of view, we are also interested in the
gauge theory on this noncommutative space.

In order to formulate the local $U(1)$ gauge theory on
the fuzzy sphere, we first have to define the differential algebra 
based on the above algebra $\CA_N$. 
We apply Connes' framework of noncommutative 
differential geometry \cite{ConnesNCG} by using 
a spectral triple ($\CA_N$,$\CH_N$, $\BD$) 
proposed recently by the authors\cite{CW97a},
where $\BD$ is the Dirac operator and 
$\CH_N$ is the corresponding Hilbert space of spinors.
We analyze the space of $1$-forms which corresponds to the
gauge potential and give the $2$-forms to define the field strength.

This paper is organized as follows. In section 2, we 
summarize the definitions of the Dirac operator, the chirality operator and 
the spectral triple.  We give a complete derivation of 
the spectrum of the Dirac operator
and discuss its properties in detail. 
Then we define the differential algebra on the fuzzy sphere.
 In section 3, the gauge field and the field strength are defined using 
this differential algebra. We examine the structure of the $U(1)$ gauge 
transformation of the charged scalar field. 
Then the corresponding invariant actions are formulated. 
Section 4 contains the discussion. We also discuss the 
commutative limit.

\section{Noncommutative Differential Algebra}

%%%

\subsection{Algebra of Fuzzy Sphere}

The algebra of the fuzzy sphere can be obtained by
quantizing the function algebra over the sphere
by using its Poisson structure. 
For this end we adopt the Berezin-Toeplitz 
quantization which gives the quantization procedure for 
a K\"ahler manifold \cite{Berezin74,Berezin75a}. 
Applying this method to the 
function algebra over the sphere we obtain
the algebra $\CA_N$. $\CA_N$ can be represented by 
operators acting on a $(N+1)$ dimensional Hilbert space $\CF_N$.
The algebra $\CA_N$ can thus be identified with the 
algebra of the complex $(N+1)\times(N+1)$ matrices.

The basic algebra to be quantized is
the function algebra $\CA_\infty$ of the square integrable 
functions over a $2$-sphere.
The basis of this algebra is given by the 
spherical harmonics $Y_{lm}$ and the multiplication of the algebra
is a usual pointwise product of functions.
The fuzzy sphere may also be introduced as an approximation of
 the function algebra over 
the sphere by taking a finite number $N$ of spherical harmonics, 
where this number $N$ is limited by the 
maximal angular momentum $\{Y_{lm}; l\leq N\}$. 
However with respect to the usual multiplication 
this set of functions does not form a closed algebra 
since the product of two spherical harmonics $Y_{lm}$
and $Y_{l'm'}$ contains $Y_{l+l',m}$. 
It is a new multiplication rule that solves the above 
described situation and 
gives a 
closed function algebra with a finite number of basis elements.
The resulting algebra $\CA_N$ is noncommutative.
We can identify
the algebra of the fuzzy sphere with the algebra of complex matrices 
$M_{N+1}(\complex)$ and thus we can consider it as a special case of 
matrix geometry \cite{DuboKern89a,DuboKern90,DuboKern90a,DuboMado91}.

The operator algebra $\CA_N$ and the Hilbert
space $\CF_N$ can be formulated keeping the symmetry 
properties under the rotation group.
We introduce a pair of creation-annihilation
 operators $\Ba^{\dagger b}, \Ba_b$ ($b=1,2$) 
which transforms as a fundamental representation under the $SU(2)$ action
of rotation.

\be
[\Ba^a,\Ba^\dagger_b]=\delta^a_b \ .\label{BosonicCommutator}
\ee  
Define the number operator by
$\BN=\Ba^\dagger_b\Ba^b \ ,
$
then the set of states $|v>$ in the Fock space associated with the 
creation-annihilation operators satisfying
\be
\BN|v>=N|v> \ ,
\ee
provides an $N+1$ dimensional Hilbert space $\CF_N$.
The orthogonal basis $|k>$ of $\CF_N$ can be defined as 
\be
|k\rangle ={1\over \sqrt{k!(N-k)!}}(\Ba^{\dagger}_1)^k
(\Ba^{\dagger}_2)^{N-k}|0\rangle\ ,
\label{Fockbasis} 
\ee
where $k=0,...,N$ and $|0\rangle$ is the vacuum.

The operator algebra $\CA_N$ acting on $\CF_N$ 
is unital and given by operators $\{\CO; [\BN,\CO]=0\}$.
The  generators of the algebra $\CA_N$ are defined by 
\be
\Bx_i=\half\alpha\sigma_i^{a}{}_b\Ba^\dagger_a\Ba^b \ ,
\ee
where the normalization factor $\alpha$ 
is a central element $[\alpha, \Bx_i]=0$ and is defined 
by the constraint
\be
\Bx_i\Bx_i={\alpha^2\over4}\BN(\BN+2)=\ell^2.\label{radius}
\ee
The above equation means that $\ell>0$ is the radius of the $2$-sphere 
and we get for $\alpha$ 
\be
\alpha={2\ell\over\sqrt{\BN(\BN+2)}}\ .\label{alpha}
\ee

The algebra of the fuzzy sphere is generated by $\Bx^i$ and the basic
relation is
\be
[\Bx_i, \Bx_j]=i\alpha\epsilon_{ijk}\Bx_k.\label{fuzzyalgebra}
\ee
On the Hilbert space $\CF_N$, $\alpha$ is constant and plays the 
role of  the "Planck constant". 
The commutative limit corresponds to 
$\alpha\rightarrow0$, i.e., $N\rightarrow\infty$.\footnote{Another possible 
choice is to take
$\alpha={2\over N}$ as in ref.\cite{Berezin74}. With this choice, 
the radius of the fuzzy sphere depends on $N$.}

Now let us consider the derivations of 
$\CA_N$. 
Among them, the derivative operator $\BL_i$ is defined by the adjoint action 
of $\Bx^i$ \cite{MadoreCQG92}
  
\be
{1\over \alpha}ad_{\Bx_i}\Ba
={1\over \alpha}[\Bx_i,\Ba] \equiv \BL_i \Ba\ , \label{defL}
\ee
where $\Ba\in \CA_N$.
These objects are the noncommutative analogue of the
Killing vector fields on the sphere, and the algebra of $\BL_i$ closes.
We obtain thus
\bea
[\BL_i,\Bx_j]=i\epsilon_{ijk}\Bx_k\ ,\qquad
[\BL_i,\BL_j]=i\epsilon_{ijk}\BL_k\ .
\eea

Finally, the integration is given by the
trace over the Hilbert space $\CF_N$.
The integration over the fuzzy sphere which corresponds to the
standard integration over the sphere in the commutative limit is defined by
\be
\langle\CO\rangle={1\over N+1}\Tr{\CO}
={1\over N+1}\sum_k \langle k|\CO|k\rangle \ .
\ee
where $\CO\in\CA_N$.

\subsection{Chirality Operator and Dirac Operator}

We introduce the spinor field $\Psi$ as an $\CA_N$-bimodule
$\Gamma\CA_N\equiv\complex^2\otimes \CA_N$, which is the noncommutative 
analogue of the space of sections of a spin bundle.
$\Psi$ is represented by 2-component spinors $\Psi=\mat{\psi^1\cr\psi^2}$
where each entry is an element of $\CA_N$ and 
we require that it  transforms
as a spinor under rotation of the sphere.

Since left multiplication and right multiplication commute, 
the $\CA_N$-bimodule can be considered as a left module over the algebra
 $\CA_N\otimes \CA_N^{\Bo}$, 
where $\CA_N^{\Bo}$ denotes the opposite algebra which is defined by:
\be
\Bx_i^{\Bo}\Bx_j^{\Bo}\equiv (\Bx_j\Bx_i)^{\Bo} \ , \ 
\Bx_i\in \CA_N \ .
\ee
The action of $\Ba ,\Bb \in\CA_N$ 
onto the $\CA_N$-bimodule
$\Psi\in\Gamma\!\CA_N$ is
\be
\Ba\Bb^{\Bo}\,\Psi\equiv \Ba\,\Psi\,\Bb \ .
\ee

We define the Dirac operator and the chirality operator 
in the algebra $\CA_N\otimes \CA_N^{\Bo}$ \cite{CW97a},
i.e. as $2\times2$ matrices the entries of which are
elements in the algebra $\CA_N\otimes \CA_N^{\Bo}$.
The construction of the Dirac operator is performed 
by the following steps:

\begin{itemize}
\item[(a)] Define a chirality operator which commutes with the elements of 
$\CA_N$ and which has a standard commutative limit.
\item[(b)] Define the Dirac operator by requiring that 
it anticommutes with the chirality operator and, 
 in the commutative limit it reproduces the standard Dirac
operator on the sphere.
\end{itemize}

Requiring the above condition (a) we obtain 
for the chirality operator \cite{CW97a} 
\be
\gamma_\chi={1\over \CN}(\sigma_i\Bx^{\Bo}_i-{\alpha\over 2}) \ .
\label{chirality}
\ee
$\CN$ is a normalization constant 
defined 
by the condition 
\be
(\gamma_\chi)^2=1 \ ,
\ee
as $\CN={\alpha \over 2}(\BN+1)$ and
$\sigma_i$ ($i=1,2,3$) are the Pauli matrices.
In the commutative limit, the operator $\Bx_i$ can be identified with
the homogeneous coordinate $x_i$ of sphere and 
the chirality operator given in eq.(5) becomes 
${1 \over\ell}\sigma_ix_i$
 which is the standard
chirality operator invariant under rotation.\cite{Jayewardena88}

The chirality operator (\ref{chirality}) 
defines a $Z_2$ grading of the differential algebra and it 
commutes with the algebra $\CA_N$.

\proposition{1}

The Dirac operator $\BD$ satisfying 
the condition (b), i.e., $\{\gamma_\chi,\BD\}=0$, is given by
\be
\BD={i\over \ell\alpha}
\gamma_\chi\epsilon_{ijk}\sigma_i\Bx_j^{\Bo}\Bx_k\ .
\label{Dirac0}
\ee

\proof See ref.\cite{CW97a}.\footnote{Note that this 
Dirac operator is different from the one given in 
ref.\cite{GrosMado95}. The difference is that the operator in 
ref.\cite{GrosMado95} contains a product of Pauli matrix and angular 
momentum operator, whereas the operator defined here 
contains a product 
of $\chi_i$ and angular momentum operator as in eq.(\ref{Dirac1}), 
i.e. it also contains 
$\Bx_i$. Consequently, the spectra are not the same.} 

\medskip
Note that this Dirac operator is selfadjoint,  
$\BD^\dagger = \BD$.

Acting with this operator on a spinor $\Psi\in\Gamma\CA_N$,
we obtain
\be
\BD\Psi={i\over \ell}\gamma_\chi\chi_i\BJ_i\Psi\ ,
\label{Dirac1}
\ee
where
\be
\BJ_i=\BL_i+\half\sigma_i \ ,\label{TotalAng}
\ee
and
\be
\chi_i\equiv\epsilon_{ijk}\Bx_j\sigma_k\ .
\ee
The action of the angular momentum operator 
on the bimodule is defined by 
\be
\BL_i\Psi\equiv {1\over\alpha}[\Bx_i,\Psi]={1\over\alpha}(\Bx_i\Psi-\Psi\Bx_i)
\ .
\ee

The second condition of (b) concerning the commutative limit of the Dirac 
operator is also satisfied. 
If we replace each operator 
$\chi_i, \BJ_i$ and $\gamma_\chi$ in eq.(\ref{Dirac1}) 
by the corresponding quantity which is obtained in 
the commutative limit, we get

\be
\BD_\infty={i\over\ell}\gamma_\chi\chi_i\BJ_i
={i\over \ell^2}(\sigma_l x_l)\epsilon_{ijk} x_i\sigma_j (i\BK_k+\half\sigma_k)
=- (i\sigma_i \BK_i+1)\ ,
\ee
where $x_i$ is the homogeneous coordinate of $S^2$ and $\BK_i$ is the 
Killing vector. 
Therefore, in the commutative limit 
this Dirac operator is equivalent to 
the standard Dirac operator.

\subsection{Spectral Triple}

In order to establish Connes' triple we have 
to identify the Hilbert space. 

The space of the fermions $\Psi\in\CA_N\otimes \complex^2$ 
defines a Hilbert space $\CH_N$ with norm 
\be
\langle \Psi|\Psi \rangle=\tr_{\CF}(\Psi^\dagger\Psi)=\sum_{\rho=1}^2
\tr_{\CF}\{(\psi^\rho)^*\psi^\rho\}\ ,
\ee 
where $\tr_{\CF}$ is the trace over the $(N+1)$ dimensional 
Hilbert space
$\CF_N$.

The dimension of the Hilbert space $\CH_N$ is $2(N+1)^2$ and the 
trace over
$\CH_N$ is the trace 
over the spin suffices and over the $(N+1)^2$ 
dimensional space of the matrices.
Since the Dirac operator is defined in the algebra 
$\CA_N\otimes\CA_N^{\Bo}$,
the trace must be taken 
for operators of the form $\Ba\Bb^{\Bo}$, 
with $\Ba,\Bb\in\CA_N$, and it is given by
\be
\tr_{\CH}\{\Ba\Bb^{\Bo}\}=\sum^{2(N+1)^2}_{K=1}
\langle \Psi_K|\Ba\Bb^{\Bo}\Psi_K\rangle
=2\tr_{\CF}\{\Ba\}\tr_{\CF}\{\Bb\} \ .\label{trace}
\ee
Here $\Psi_K$ is an appropriate basis in $\CH_N$ labeled by
an integer $K\in\{1,...,2(N+1)^2\}$. The factor $2$ 
on the r.h.s. comes from the trace over the spin suffices.

To examine the structure of the Hilbert space we compute the 
spectrum $\lambda_j$ of the Dirac operator:
\be
\BD^2\Psi_{jm}=\lambda_j^2\Psi_{jm}\ .
\ee
$\Psi_{jm}$ is a state with total angular momentum $j$, 
$\BJ^2\Psi_{jm}=j(j+1)\Psi_{jm}$ and $\BJ_3\Psi_{jm}=m\Psi_{jm}$
is the $\Bx_3$ component of the total angular momentum 
operator $\BJ_i$ in eq.(\ref{TotalAng}).
$j$ and $m$ are half integers and run $\half\leq j\leq N+\half$ and 
$-j\leq m\leq j$.

\proposition{2}

The spectrum of the Dirac operator is given by 
\be
\lambda^2_j=(j+\half)^2[1
+{1-(j+\half)^2\over N(N+2)}]\ .\label{spectrum}
\ee

\proof
\bea
{\ell^2\over\alpha^2}\BD^2
&=&(\epsilon_{ijk}\sigma^i\BX^j\BY^k)
(\epsilon_{i'j'k'}\sigma^{i'}\BX^{j'}\BY^{k'})\cr
&=&\BX^2\BY^2-(\BX\BY)[(\BX\BY)+1+(\BX\sigma)+(\BY\sigma)]\ ,
\eea
where $\BX_i={1\over\alpha}\Bx_i$, $\BY_i=-{1\over\alpha}\Bx^\Bo_i$
and $(\BX\BY)=\sum_i\BX_i\BY_i$.
Using the relations 
\bea
\BL_i=\BX_i+\BY_i\and\BJ_i=\BL_i+\half\sigma_i\ ,
\eea
we obtain $(\BX\BY)=\half[\BL^2-\BX^2-\BY^2]$ and 
$(\sigma\BX)+(\sigma\BY)=\BJ^2-\BL^2-{3\over4}$.

In order to evaluate the spectrum we use the representation
of the spinor and substitute
\be
\BJ^2=j(j+1)\and \BL^2=(j+s)(j+s+1)\ ,
\ee
where $j\leq N+\half $ is a half integer and $s=\pm\half$. 
With this value we get
\bea
(\BX\BY)
%&=&\half[(j+s)(j+s+1)-\BX^2-\BY^2]\cr
&=&\half[j(j+1)+s(2j+1)+{1\over4}-\BX^2-\BY^2]\cr
(\sigma\BX)+(\sigma\BY)\ 
%&=&j(j+1)-(j+s)(j+s+1)-{3\over4}\cr
&=&-s(2j+1)-1\ .
\eea
Thus, the eigenvalue is
\be
{\ell^2\over\alpha^2}\lambda_j^2
=-{1\over4}(j+\half)^2\Big[
(j+\half)^2-2(\BX^2+\BY^2)-1\Big]
-{1\over4}(\BX^2-\BY^2)^2\ .
\ee
If we substitute $\BX^2=\BY^2={N\over2}({N\over2}+1)$ we obtain
the relation (\ref{spectrum}).
\qed

This spectrum 
coincides with the classical spectrum of the Dirac operator
in the limit $N\rightarrow \infty$. 
For finite $N$, it contains zeromodes.
When the angular momentum takes its maximal value 
we see that $\lambda_{N+\half}=0$.  This happens since there is no
chiral pair for the spin $N+\half$ state and therefore this part must be 
a zeromode for consistency. We can also confirm this property by 
computing: $\tr_{\CH}(\gamma_\chi)=2(N+1)$. 
Since these zeromodes have no classical analogue, one way 
to treat them is to project them out from the Hilbert space. 
On the other hand, 
the contribution of the zeromodes in the integration 
is of order ${1\over N}$ and thus their contribution vanishes 
in the limit $N\rightarrow \infty$. 
Therefore, considering the differential algebra on 
the fuzzy sphere as a kind of regularization of the
 differential algebra on the sphere, 
it is sufficient to take the full 
Hilbert space $\CH_N$.

In this way we obtain Connes' triple $(\CA_N, \BD, \CH_N)$.
We thus can apply the construction of the differential algebra. 

\subsection{Differential Algebra}

In this section we construct the differential algebra associated 
with $(\CA_N, \BD, \CH_N)$ by using Connes' method \cite{ConnesNCG}. 
See also \cite{ChamFroh93}.

We define the universal differential algebra $\Omega^*(\CA_N)$ 
over $\CA_N$. An element $\omega\in\Omega^*(\CA_N)$
 is in general given by
\be
\omega=\sum_{\lambda\in \CI} \Ba_\lambda^{(0)}
d\Ba_\lambda^{(1)}d\Ba_\lambda^{(2)}\cdots d\Ba_\lambda^{(p)}\ ,
\ee
where $p$ is an integer, 
$\Ba_\lambda^{(k)}\in \CA_N \ (k=0\cdots p)$ and $\CI$ is an appropriate
 set labeling the elements. $d\Ba$ is a symbol defined by the 
operation of the differential $d$ 
on $\Ba\in\CA_N$, which satisfies Leibnitz rule
$d(\Ba\Bb)=(d\Ba)\Bb+\Ba (d\Bb)$ for $\Ba,\Bb\in\CA_N$, and
$d\id=0$ for the identity $\id\in\CA_N$.  We also require
$(d\Ba)^*=-d\Ba^*$. The Leibnitz rule provides a natural product among 
the elements in $\Omega^*(\CA_N)$ and the differential $d$ on 
$\Omega^*(\CA_N)$ is defined by
\be
d(\sum_{\lambda\in \CI} \Ba_\lambda^{(0)}
d\Ba_\lambda^{(1)}d\Ba_\lambda^{(2)}\cdots d\Ba_\lambda^{(p)})
=\sum_{\lambda\in \CI} d\Ba_\lambda^{(0)}
d\Ba_\lambda^{(1)}d\Ba_\lambda^{(2)}\cdots d\Ba_\lambda^{(p)}\ .
\ee
Then, it follows  $d^2\omega=0$ and the graded 
Leibnitz rule.

In order to define the $p$-forms as operators on $\CH_N$, 
a representation $\pi$ is defined by
\be
\pi(\sum_{\lambda\in \CI} \Ba_\lambda^{(0)}
d\Ba_\lambda^{(1)}d\Ba_\lambda^{(2)}
\cdots d\Ba_\lambda^{(p)})
=\sum_{\lambda\in \CI} \Ba_\lambda^{(0)}
[\BD,\Ba_\lambda^{(1)}][\BD,\Ba_\lambda^{(2)}]
\cdots [\BD,\Ba_\lambda^{(p)}]\ .
\ee
Recall that $\CA_N$ is defined as an algebra of operators
in $\CH_N$.
Then the graded differential algebra is defined by
\be
\Omega^*_{\BD}(\CA_N)=\Omega^*(\CA_N)/\BJ\ ,
\ee
where $\BJ=ker\pi+d\, ker\pi$ is the differential ideal of 
$\Omega^*(\CA_N)$.

In order to establish the differential calculus on
the fuzzy sphere, we have to examine the structure of the
differential kernel $\BJ$. 
For this we denote the kernel of each level as
\be
ker\pi^{(p)}\equiv \Omega^p(\CA_N)\cap ker\pi\ ,
\ee
then the differential kernel $\BJ^{(p)}$ for the $p$-form is
\be
\BJ^{(p)}=ker\pi^{(p)}+d\,ker\pi^{(p-1)}\ .
\ee

Since the elements of the algebra $\CA_N$ are defined as
operators in $\CH_N$,  $ker\pi^{(0)}=\{0\}$, 
i.e., $\BJ^{(0)}=\{0\}$. 
It means that $\Omega_{\BD}^0(\CA_N)=\CA_N$.
The differential kernel of the $1$-form is
$\BJ^{(1)}=ker\,\pi^{(1)}+d\,ker\pi^{(0)}=ker\,\pi^{(1)}$,
and thus for any element $\Ba \in \CA_N$ 
the derivative is defined by 
\be
\pi(d\Ba)=[\BD,\Ba]\ .
\ee
The space of $1$-forms $\omega\in\Omega_{\BD}^1(\CA_N)$ can be 
identified with the operators 
$\pi(\omega)$ in 
$\CH_N$:
\be
\pi(\Omega_{\BD}^1(\CA_N))=\{ \pi(\omega) |\ \pi(\omega)=
\sum_{\lambda\in\CI} \Ba_\lambda[\BD,\Bb_\lambda]\ ;
\ \Ba_\lambda,\Bb_\lambda\in \CA_N\} \ .\label{oneform}
\ee
Thus, with the above identification, the exterior derivative 
$d$ defines a map:
\be
d\quad :\quad \CA_N\ \rightarrow\  
M_2({\complex})\otimes(\CA_N\otimes \CA_N^{\Bo})\ ,
\ee
where $M_2({\complex})$ is the algebra of 
$2\times2$ complex matrices.

Using the definition of the Dirac operator (\ref{Dirac0}),
a $1$-form is expressed as follows:
Take a $1$-form $\pi(\omega)\in\pi(\Omega^1_{\BD}(\CA_N))$ 
in eq.(\ref{oneform}). 
Using eq.(\ref{Dirac0}) we obtain 

\be
\pi(\omega)=\sum_\lambda {i\over \ell\alpha}
\gamma_\chi\epsilon_{ijk}
\sigma_i\Bx_j^{\Bo}\Ba_\lambda[\Bx_k, \Bb_\lambda]
={i\over\ell}\gamma_\chi\chi^{\Bo}_k\omega_k\ ,
\label{general1form}
\ee
where
\be
\chi^{\Bo}_k\equiv \epsilon_{ijk}\sigma_i\Bx_j^{\Bo}\ ,
\ee
and the components $\omega_k$ of $\pi(\omega)$ 
can be rewritten by 
using the definition (\ref{defL}) of $\BL$ as:
\be
\omega_k
\equiv {1\over \alpha}\sum_\lambda\Ba_\lambda[\Bx_k, \Bb_\lambda]
=\sum_\lambda\Ba_\lambda(\BL_k\Bb_\lambda)\ .\label{oneform1}
\ee
Here, 
$\omega_k\in\CA_N$ may be considered as the component of a 
vector field.

\medskip%one-form is in the appendix

In order to write the gauge field action,
 we have to define the $2$-form. 
A $2$-form $\eta\in \Omega_\BD^2(\CA_N)$ 
can be given in general as
\bea
\pi(\eta)&=&\sum_\lambda \Ba_\lambda^{(1)}[\BD,\Ba^{(2)}_\lambda]
[\BD,\Ba^{(3)}_\lambda] \cr
&=&{1\over\ell^2}\chi^o_i\chi^o_j\sum_\lambda \Ba_\lambda^{(1)}(\BL_i\Ba^{(2)}_\lambda)
(\BL_j\Ba^{(3)}_\lambda) \ .\label{2form}
\eea
where $\Ba^{(i)}_\lambda\in \CA_N$.
Since the $2$-form in eq.(\ref{2form}) is defined 
 up to the differential kernel
 $\pi(dker\pi^{(1)})$,
$\pi(\eta)$ contains redundant components.

 Note that 
when we perform the calculation, we do not use
 the $\Omega_{\BD}^2(\CA_N)$, but 
its representation $\pi(\Omega^2_{\BD}(\CA_N))$,
thus it is sufficient to compute $\pi(d\,ker\pi^{(1)})$, since
$\pi(\Omega_{\BD}^*(\CA_N))$ is isomorphic to
$\pi(\Omega^*(\CA_N))/\pi(d\,ker\pi)$.

The 
nontrivial contribution of $\pi(d ker\pi^{(1)})$
is proportional to the traceless part of the symmetric product 
$\chi^o_{\{i}\chi^o_{j\}}$ as we shall see in the following.

The exterior derivative of a general $1$-form $\omega$
defined in eq.(\ref{oneform}) is
\be
\pi(d\omega)
=\sum_\lambda [\BD,\Ba_\lambda][\BD, \Bb_\lambda]\ .\label{dof1form}
\ee
Using the Dirac operator we obtain

\bea
\pi(d\omega)
&=&{1\over\ell^2}\sum_\lambda\chi_i^{\Bo}\chi_{i'}^{\Bo}
\Big[\BL_i(\Ba_\lambda\BL_{i'}\Bb_\lambda)
-i \half\epsilon_{ii'k}\Ba_\lambda\BL_k\Bb_\lambda
+{2\over3\alpha}\delta_{i,i'}(\Ba_\lambda\BL_i\Bb_\lambda)\Bx_i\cr
&&\qquad-\Ba_\lambda[\half\{\BL_i,\BL_{i'}\}-{1\over3}
\delta_{ii'}\BL^2]\Bb_\lambda\Big]
\label{kernel2}
\eea
The first three terms vanish for
 $\omega\in ker\pi^{(1)}$.  Only the last term gives 
 a nontrivial contribution for
the differential kernel and thus 
$\pi(d\, ker\pi^{(1)})$ is proportional to
 the symmetric traceless product of 
$\chi^o_i\chi^o_{i'}$.

The proof of the existence of the nontrivial 1-form 
kernels which contribute to 
$d ker\pi^{(1)}$ 
is given in the appendix. Using the
explicit expression $\omega_{p,q}=\Bx_A^p d \Bx_A^q$ of 
$ker\pi^{(1)}$ obtained in the appendix (see eq.(\ref{kernel_omega}))
we compute $\pi( d\ ker \pi^{(1)})$. We find 
that 
$d\omega_{p,q}$ gives an element of $d \,ker\pi^{(1)}$:

\proposition{3}
$\pi(d\omega_{p,q})\not=0$, for $p+q=N+2$ and $p,q>1$.

\proof
\bea
\pi(d \omega_{p,q})
&=& [D,\Bx_+^p][D,\Bx_+^q]\cr
&=&
{-1\over \ell^2}\gamma_\chi \chi_+^o \gamma_\chi \chi_+^o [-2p\Bx_+^{p-1}
(\Bx_3+{\alpha\over 2}(p-1))][-2q\Bx_+^{q-1}(\Bx_3+{\alpha\over 2}(q-1))]\cr
&=& {1\over \ell^2}\chi_+^o\chi_+^o 4pq 
\Bx_+^{p+q-2}\Big(\Bx_3^2+\Bx_3({\alpha\over 2}p+
{3\alpha\over 2}q-2\alpha) \cr
&&~~~~~~~~~~~~~~~ ~~~~~~~~~~~~~~~~~ ~~~~~~~~~~~~
 +{\alpha^2\over 4}(p+2q-3)(q-1)\Big)
\eea
Using the identity $\Bx_+\Bx_-=\ell^2+\alpha\Bx_3-\Bx_3^2$,
this expression can be simplified to 

\be
\pi(d \omega_{p,q})=4pq{1\over \ell^2}\chi_+^o\chi_+^o\Bx_+^{p+q-2}[A(q)\Bx_3+B(q)]
\ee
where 
\be
A(q)={\alpha\over 2}p+{3\alpha\over 2}q-\alpha\ ,
\ee
\be
B(q)=\ell^2+{\alpha^2\over 4}(p+2q-3)(q-1)-\Bx_+\Bx_- \ .
\ee
This means that $\pi(d \omega_{p,q})$ does not vanish
for $p+q=N+2$ although $\omega_{p,q}\in ker\pi^{(1)}$ for
$p+q=N+2$.
\qed

The result of the proposition 3 and the corresponding contribution from 
the kernels of the other directions 
show that 
there exist nontrivial differential kernel elements 
$\pi(d\, ker\pi^{(1)})$
proportional to the symmetric traceless product of $\chi^o_i\chi^o_j$.

With this result we can prove the following proposition.

\proposition{4}
\be
\pi(d\, ker\pi^{(1)})=\{\Lambda|\Lambda={1\over\ell^2}
\chi_i^{\Bo}\chi_{j}^{\Bo}\Ba_{ij}\where\Ba_{ij}\in\CA_N,
\Ba_{ij}=\Ba_{ji}\and \sum_{i=1}^3\Ba_{ii}=0\}\ .
\ee

\proof
Using Proposition 3, we obtain a nontrivial element
by multiplying $\Ba'_\lambda,\Bb'_\lambda\in\CA_N$ and
\bea
\pi\Big(d (\sum_{\lambda}\Ba'_\lambda\omega_{p,q}\Bb'_\lambda)\Big)
&=&\sum_{\lambda}\pi\Big(\Ba'_\lambda(d \omega_{p,q})\Bb'_\lambda\Big)\cr
&=&\sum_{\lambda} {1\over\ell^2}\chi^o_+\chi^o_+
\Ba'_\lambda(\BL_-\Bx_+^p)(\BL_-\Bx_+^q)\Bb'_\lambda\cr
&=&
4pq{1\over \ell^2}
\chi_+^o\chi_+^o\sum_{\lambda} 
\Ba'_\lambda\Bx^N_+[A(q)\Bx_3+B(q)]\Bb'_\lambda
\eea
where we have used $p+q=N+2$. Choosing appropriate elements 
$\Ba'_\lambda,\Bb'_\lambda\in\CA_N$, the factor 
$\sum_{\lambda}\Ba'_\lambda\Bx^N_+[A(q)\Bx_3+B(q)]\Bb'_\lambda$
 can become any element in $\CA_N$.
We have six independent directions for $\omega_{p,q}$ and 
combining the results from them we get 
the traceless symmetric combinations of
suffices $i,i'$ in $\chi^o_{i}\chi^o_{i'}$.
\qed

Identifying the $\Omega^2_{\BD}(\CA_N)$ with its representation
$\pi(\Omega^2_{\BD}(\CA_N))$, 
a general $2$-form $\widetilde{\eta}\in \pi(\Omega^2_{\BD}(\CA_N))$ is given by
\be
\widetilde{\eta}=\sum_\lambda \Ba^{(1)}_\lambda[\BD,\Ba^{(2)}_\lambda][\BD,\Ba^{(3)}_\lambda]\ ,
\ee
where $\Ba^{(i)}_\lambda\in\CA_N$ up to $\pi(d\,ker\pi^{(1)})$.

Combining eqs.(\ref{dof1form}) and (\ref{kernel2})
we can compute the operation of the derivative $d$
on a general $1$-form in eq.(\ref{oneform}) and we obtain
\be
\pi(d\omega)={1\over 2\ell^2}
\chi_k^{\Bo}\chi_{k'}^{\Bo}
\Big(\{\BL_k\omega_{k'}-\BL_{k'}\omega_k\}
- i\epsilon_{kk'k''} \omega_{k''}
+\delta_{kk'}{2\over3\alpha}[\Bx_i\omega_{i}+\omega_{i}\Bx_i]\Big)\ ,
\ee
where we have used the definition of the components $\omega_k$
 in eq.(\ref{general1form}).

Since the trace part does not 
belong to the differential kernel, the last term in the above 
equation is not removed by dividing differential kernels.
We continue here our construction of the gauge field action 
with this definition of the differential algebra
and
we shall obtain a kind of mass term in the gauge theory.
The commutative limit $\alpha\rightarrow0$ becomes
singular, as can be seen from eq.(\ref{general1form}). 
However, as we discuss in the following, 
we can still interpret 
 the resulting theory as a regularization
of the corresponding commutative theory.

An alternative strategy to the one taken here would be 
to restrict the above defined 2-form. With the above $2$-form 
as it stands the naive commutative limit
does not give the standard differential calculus. 
One possibility to handle this situation is to use the property of
the trace: $\chi^o_i\chi^o_i=2\CN^2-\alpha\CN\gamma_\chi$.
It turns out 
that the trace part $J_T$ is an ideal of the $\pi(\Omega^2(\CA_N))$.
Furthermore, in each $p$-form space $\pi(\Omega^p(\CA_N))$,
the set $J_T\pi(\Omega^{p-2}(\CA_N))\cup\pi(\Omega^{p-2}(\CA_N))J_T$ 
is an ideal and thus there is a possibility to 
divide the differential algebra so that we can take the 
commutative limit and obtain the standard differential 
calculus. This procedure will be discussed elsewhere.

\section{$U(1)$ Gauge Field Theory}

\subsection{Vector Field}

Using the geometric notions defined in the previous sections,
we formulate the $U(1)$ gauge theory on the fuzzy sphere. 
We identify the differential algebra 
$\Omega_{\BD}^*(\CA_N)$
with its representation $\pi(\Omega_{\BD}^*(\CA_N))$ and
do not write the map $\pi$ explicitly.

First, to formulate the gauge field theory 
we define the real vector field $\BA$ which is a $1$-form on the fuzzy sphere.
We impose the reality condition for this $1$-form by

\be
\BA^\dagger=\BA \label{Hermiticity}\ .
\ee
Using the general definition of a $1$-form, $\BA$ can be 
written  as\footnote{The hermiticity condition
requires the form
$
\BA=\sum_\rho \Ba_\rho[\BD,\Bb_\rho]
+\Bb_\rho^*[\BD,\Ba_\rho^*]
-\half[\BD,\Ba_\rho\Bb_\rho+\Bb^*_\rho\Ba^*_\rho].
$ This can be again written in the form (\ref{potential}).}

\be
\BA=\sum_\lambda \Ba_\lambda[\BD, \Bb_\lambda]\ ,
\label{potential}
\ee
where $\Ba_\lambda,\Bb_\lambda\in\CA_N$ 
are appropriate elements.
According to the general discussion about $1$-forms in 
the previous section
 we can write
\be
\BA
={i\over\ell}\gamma_\chi\chi^{\Bo}_k\BA_k \ ,
\ee
where $\BA_k$ is the component field of $\BA$ given by
\be
\BA_k
=\sum_\lambda\Ba_\lambda(\BL_k\Bb_\lambda)\ . \label{gf}
\ee
For the component field the reality condition gives
\be
\BA_k^*=\BA_k\ .
\ee
Thus each component of the gauge field is represented 
by an $(N+1)\times(N+1)$ hermitian matrix.

Note that, in the commutative case, 
the $1$-form satisfies the constraint $\Bx_i\BA_i=0$, which shows 
the reduction of the degrees of freedom. However
in the noncommutative case the $1$-form defined 
by the equation (\ref{gf})
does not satisfy the similar
constraint on $\BA_k$ in general. 
Further discussion on the treatment of 
this property is given in section 4.

In the remaining part, let us push forward the construction of the 
gauge theory on the noncommutative sphere. 
In the commutative case, we obtain the field strength of the
$U(1)$ gauge theory by taking the exterior derivative of the
$1$-form.  In the noncommutative case, the exterior derivative 
gives
\be
d\BA=\sum_\lambda [\BD,\Ba_\lambda][\BD, \Bb_\lambda]\ .
\ee
Applying the result of the previous section we obtain
\be
d\BA={i\over 2\ell^2}
\chi_k^{\Bo}\chi_{k'}^{\Bo}
\BF_{kk'}\ ,
\ee
with
\be
\BF_{kk'}=
-i\{\BL_k\BA_{k'}-\BL_{k'}\BA_k\}-\epsilon_{kk'k''} \BA_{k''}-
i\delta_{kk'}{2\over3\alpha}[\BA_i\Bx_{i}+\Bx_{i}\BA_i]  \ .
\label{FieldStrength}
\ee

We can show that the above $\BF_{kk'}$ corresponds to the
 field strength for the 
abelian gauge field in the commutative limit.
To see this, we use the following correspondence
which holds in the commutative limit:
\be
\BA_k=\BK_k^\mu A_\mu \and \BL_k=i\BK_k\ . \label{identifications}
\ee
Here $A_\mu$ is a gauge field and  
$\BK_k^\mu$ ($k=1,2,3$, $\mu=1,2$) is the Killing vector on the
sphere with appropriate coordinates $\rho^\mu$, and $\BK_k=\BK_k^\mu\p_\mu$.
With the above identification we get
\be
\BF_{kk'}
=\BK^\mu_k\BK^\nu_{k'}F_{\mu\nu}\ ,
\ee
where $F_{\mu\nu}=\p_\mu A_\nu-\p_\nu A_\mu$. 
Here we have used the relation $\BA_i\Bx_i=0$ which holds in the 
commutative case.

In the noncommutative case, however, the exterior derivative of the
$1$-form $d\BA$ does not give the field strength.

\medskip

\subsection{$U(1)$ Gauge Transformation}

For the formulation of 
the $U(1)$ gauge theory on the fuzzy sphere, let us consider 
the $U(1)$ gauge transformation of a charged scalar field,
i.e., a complex scalar field \cite{CW97a}.
The algebraic object corresponding to the complex scalar field 
on the fuzzy sphere is the $\CA_N$-bimodule $\Phi\in \CA_N$.
Its action is given by 

\be
\BS={1 \over 2(N+1)^2}\tr_{\CH}\{(d\Phi)^\dagger d\Phi\} \ .
\ee
Apparently, the above action is invariant under global $U(1)$ 
transformation of the phase 
\be
\Phi'=e^{i\phi}\Phi \ .
\ee 

Following the standard approach, the local $U(1)$ gauge 
transformation can be defined if we let the phase $e^{i\phi}$ 
be a function on the fuzzy sphere. In the present algebraic formulation 
this means we multiply an element $\Bu\in\CA_N$ on the field $\Phi$,
where unitarity is implemented by
\be
\Bu^*\Bu=1 \quad .\label{unitarity}
\ee

When we generalize the transformation, we may take either left or right 
multiplication of $\Bu$ on the field $\Phi$ 
due to the ordering ambiguity. 
Here we take the left multiplication as the $U(1)$ gauge 
transformation for $\Phi$:
\be
\Phi'=\Bu\Phi \ .
\ee
The transformation of the conjugate field $\overline{\Phi}=\Phi^*$ is
given by $\overline{\Phi}'=\overline{\Phi} \Bu^*$.

Since the algebra $\CA_N$ is isomorphic 
to the algebra of $(N+1)\times(N+1)$
matrices, the condition (\ref{unitarity})
shows that, as a matrix, $\Bu$ is an element of $U(N+1)$.
In other words, the local $U(1)$ gauge transformation 
on the fuzzy sphere in matrix representation is defined as 
the left $U(N+1)$ transformation.

Therefore, we define the covariant derivative $\nabla_{\BA}$ as
\be
\nabla_{\BA}\Phi
=d \Phi+\BA\Phi \ .
\ee
Then the gauge transformation of the gauge field can be defined
by requiring the covariance of $\nabla_{\BA}\Phi$:
\be
\nabla_{\BA'}(\Bu\Phi)=\Bu\nabla_{\BA}\Phi\ .
\ee
This defines the standard form of the gauge transformation 
\be
\BA'=\Bu d\Bu^*+\Bu\BA \Bu^*\ .
\label{GaugeTransformation}
\ee
In components it reads
\be
\BA'_k=\Bu(\BL_k \Bu^*)+\Bu\BA_k\Bu^*\ .
\label{GaugeTransformationC}
\ee

The above transformation keeps the hermiticity condition 
(\ref{Hermiticity}) and may be interpreted as the transformation 
of the $U(N+1)$ gauge theory on a one-point space,
and thus the covariant field strength is given by the standard 
curvature form \cite{ConnesNCG}
\be
\Theta=d\BA+\BA\BA \ .
\ee

In components the curvature $2$-form is 
\be
\Theta={-i\over2\ell^2}\chi^\Bo_k\chi^\Bo_{k'}\Theta_{kk'}\ , 
\label{Curvature}
\ee
where the component of the
field strength is 

\bea
\Theta_{kk'}
&=&i\{\BL_k\BA_{k'}-\BL_{k'}\BA_k\}+\epsilon_{kk'k''} \BA_{k''}
+i[\BA_k,\BA_{k'}]\cr
&&+{2i\over3\alpha}\delta_{kk'}[\BA_i\Bx_i+\Bx_i\BA_i
+\alpha\BA_i\BA_i]
\ .\label{NFieldStrength}
\eea

\subsection{The Action of Gauge Field and Matter}

With the above results we define the 
noncommutative analogue of 
the gauge invariant action.

The action of the charged scalar is 
\be
\BS_M={1 \over 2(N+1)^2}\tr_{\CH}\{(\nabla_{\BA}\Phi)^\dagger
 \nabla_{\BA}\Phi\} \ .\label{MatterAction}
\ee

The action of the gauge field is given by
\be
\BS_G\equiv{1\over 2(N+1)^2}\tr_{\CH}\{\Theta^2\}\ .
\label{GaugeAction}
\ee
Both actions are invariant under local $U(1)$ gauge transformation.
Thus, combining these two actions,
we obtain the action of the $U(1)$ gauge theory with
scalar matter on the fuzzy sphere. 
Note that we may introduce the 
gauge coupling constant $g$ 
by rescaling the gauge field $\BA$ to $g\BA$.

In order to see the detailed structure of the above actions, 
we take a
part of the trace.  We perform the trace relating to the opposite 
algebra and
the spin suffices. Then we obtain the action which contains only the
fields $\BA$ and $\Phi$ 
 and the trace of this action is taken over the Hilbert space
 $\CF_N$.

Then the matter action (\ref{MatterAction}) can be reduced as
\be
\BS_M={2 \over 3 (N+1)}\tr_{\CF}
\{(\BL_i\Phi+\BA_i\Phi)^*(\BL_i\Phi+\BA_i\Phi)\}\ .
\label{ReducedMA}
\ee

Similarly, the gauge field action (\ref{GaugeAction}) is reduced to
\be
\BS_G=
{C_A\over(N+1)}\tr_{\CF}\{\Theta^A_{ii'}\Theta^A_{ii'}\}+
{C_S\over(N+1)}\tr_{\CF}\{\Theta^S_{ii'}\Theta^S_{ii'}\}
\ ,\label{ReducedGA}
\ee
where 
\bea
C_A&=&{\CN^2\over2\ell^2}
\Big\{ {-\alpha^2\over3\CN^2}
+{2\over3}\Big\}\ ,\cr
C_S&=&1+{1 \over N(N+2)}
\eea
and $\Theta^S_{ij}$ ($\Theta^A$) is the (anti)symmetric part of 
the field strength given in eq.(\ref{NFieldStrength}).

Since the trace over the Hilbert space $\CF_N$ corresponds to the 
volume integration in the commutative limit, the actions  $\BS_G$
 and $\BS_M$ given in eqs.(\ref{ReducedMA}),
and (\ref{ReducedGA}) respectively,
 should correspond to the standard action on the 
sphere in the limit $N\rightarrow\infty$.

Apparently the $\Theta^S$ in the gauge action does not
have a classical correspondence. Furthermore, as we see below 
this term is singular
in the naive $N\rightarrow\infty$ limit. This is unavoidable since our 
differential algebra is singular in this limit.  

However, under certain conditions we may consider the above action 
as a regularized theory of the commutative case as follows:
The symmetric part of the action is
\be
(\Theta^S)^2\sim {1\over\alpha^2}[(\BA_i\Bx_i+\Bx_i\BA_i)
+\alpha\BA_i\BA_i]^2\ .
\ee
The above combination is gauge invariant under 
the gauge transformation given in eq.(\ref{GaugeTransformation}).
This term can be understood as 
the gauge invariant mass term of the radial component of the gauge field.
Thus, physically we can understand the effect of the symmetric part as follows: 
When we consider the quantization of the above regularized
theory using the path integral which respects the gauge symmetry, 
then in the $\alpha\rightarrow0$ limit the symmetric term
behaves like a (gauge invariant) delta function which drops the
radial component. 

Furthermore, from the point of view of gauge theory it is not
necessary 
to take $\Theta^2$ as an action. Instead, we can simply take any linear 
combination of the gauge invariant terms. This means that we can 
 take $C_A$ and $C_S$ as independent parameters.

Thus, we obtain in general the following action for the gauge field.
\be
S={1\over(N+1)}\tr_\CF\{C_1 \BG_{kk'}\BG_{kk'}+C_2\BG'^2\}
\ee
where $C_1$ and $C_2$ are c-numbers and
\bea
\BG_{kk'}&=&i\BL_k\BA_{k'}-i\BL_{k'}\BA_k
+\epsilon_{kk'k''} \BA'_{k''}+i[\BA_k,\BA_{k'}]\cr
\BG' &=&
\Bx_i\BA_i+\BA_i\Bx_i+\alpha\BA_i\BA_i\ .
\label{GeneralGaugeAction}
\eea
The above action (\ref{ReducedGA}) is a special case of the general form 
given here.

\section{Discussions and Conclusion}

In this paper we have formulated 
the $U(1)$ gauge theory on the 
fuzzy sphere, following Connes' framework of noncommutative 
differential geometry. 
The differential algebra
on the fuzzy sphere has been constructed 
by applying the chirality operator and 
Dirac operator proposed in ref.\cite{CW97a}. This chirality operator 
anticommutes with the 
Dirac operator and the structure of the differential algebra 
becomes simple. 
Then we analyzed the structure of
the $1$-forms and $2$-forms which are necessary to 
construct the gauge field action.
In ref.\cite{CW97a}, the
action of a complex scalar field on the fuzzy sphere which 
is invariant under 
the global $U(1)$ transformation of the phase of the complex scalar field has been formulated.
Here, the local $U(1)$ gauge transformation on the fuzzy sphere 
is introduced by making the global phase transformation
into a local transformation, i.e. the phase becomes 
a function over the fuzzy sphere.
By construction, a function 
over the fuzzy sphere is simply given 
by elements of the algebra $\CA_N$.  Thus, the local $U(1)$ 
gauge transformation is defined by multiplication of an 
element $\Bu\in\CA_N$, satisfying unitarity $\Bu^*\Bu=1$.

Since the algebra $\CA_N$ is noncommutative,
there is an ambiguity of operator ordering 
when replacing the global phase by the algebra elements $\Bu$.
We have chosen here the left multiplication. 
Thus, when we represent the algebra $\CA_N$ by matrices, 
the local $U(1)$ gauge transformation is identified with 
the left transformation by a unitary $(N+1)\times(N+1)$ matrix.
Therefore, the gauge field action is analogous to the Yang-Mills action.

Once we know the Dirac operator, the construction of the differential 
calculus is rather straightforward, however, as we have seen 
when defining the $1$-forms, their components $\BA_i$ do not satisfy 
$\BA_i\Bx^i=0$ in general. In the commutative case this relation holds since 
the Killing vector is perpendicular to the normal direction of the sphere. 
However in the noncommutative case $\Bx_i\BL^i$ is not necessarily zero. 
Since the relation \footnote{This relation follows from 
definition (\ref{gf}).}
\be
\Bx_i\BA_i+\BA_i\Bx_i={1\over \alpha}[\Bx_i,\Ba][\Bx_i,\Bb] \ ,
\label{radialpart}
\ee
holds,
this property is related with the trace part of the 
$2$-form as follows:

As we have seen in the construction of the $2$-forms performed here, 
the differential kernel $\pi^{(2)}$ does not contain a 
trace part, i.e., the part proportional to 
$\chi^{o}_i\chi^{o}_i$. 
In the course of deriving eq.(\ref{kernel2}) we get $[\Bx_i,\Ba][\Bx_i,\Bb]$
 as a coefficient of $\chi^{o}_i\chi^{o}_i$. Up to the kernel condition
this product of commutators is equivalent to $\Ba\BL^2\Bb$.
The reason why 
the trace part drops from the differential kernel is due to the 
relation  
$
\Ba\BL^2\Bb=-{2 \over\alpha}\Ba(\BL_i\Bb)\Bx_i \ .
$ 
This relation is a direct consequence of the condition 
that $\ell^2$ is central.
This type of problem relating to the reduction of degrees of freedom 
as well as to the structure of the differential kernel 
is a rather general feature 
when defining the differential forms 
by the adjoint action $\BL_i$. 
\footnote{The structure of the Dirac operator depends on the choice of the 
fermion, but on the other hand if the Dirac operator 
has the form $\theta^i{\Bx_i}$, and if $\theta^i$ commutes with $\Bx_i$,
 where $\theta^i\sim\gamma_\chi\chi^o_i$ 
in our case,
then the derivative 
$d$ is always given by $d a=\theta^i(\BL_i a)$ with $\BL_i$ being the 
adjoint action. }

Thus, in the noncommutative case the construction gives 1-forms which 
have three independent components. 
One possibility 
to drop the trace part (which is proportional to the third component) in the 
present approach has been indicated in 
section 2.4.

On the other hand, although the $2$-form is singular 
in the $N\rightarrow\infty$ limit, 
the action given in eq.(\ref{ReducedGA}) 
still allows the interpretation as a regularized theory
of the gauge theory on the sphere.

It is easy to check that both terms in the action eq.(\ref{ReducedGA}) 
are invariant under the gauge transformation (\ref{GaugeTransformationC}).
Thus, the most general gauge action can be written as in
eq.(\ref{GeneralGaugeAction}). The first term corresponds to the standard gauge 
action in the commutative limit. This term is usually taken as the action 
for the gauge field in the fuzzy sphere. The second term 
approaches simply $(2\Bx_i\BA_i)^2$ in this limit.

As we mentioned, the symmetric part of the
action can be understood as a gauge invariant mass for the radial 
component of the
gauge field.
Furthermore, in the action (\ref{GaugeAction}),
 this mass is diverging in the limit 
of $N\rightarrow \infty$ and can be treated 
as delta function constraint
under the path integral. Thus by taking a limit which 
respects the gauge symmetry, 
 the freedom corresponding to
$\Bx_i\BA_i+\BA_i\Bx_i+\alpha\BA^2$ is freezed 
and thus effectively drops from
the theory. Since in this limit this procedure is equivalent to the
constraint $\Bx_i\BA_i=0$, it reduces the freedom of the vector 
potential in the commutative theory properly.

From the point of view of constructing 
a gauge theory on the fuzzy sphere, we have an even 
simpler choice to treat the degrees of freedom of the theory. 
If we require only the gauge invariance under the
gauge transformation (\ref{GaugeTransformation}), 
we can take the symmetric term as a constraint
for the gauge field from the begining on.
Then the action contains only the antisymmetric part, i.e.,
$C_2=0$ in eq.(\ref{GeneralGaugeAction}) and 
the gauge field is constrained by
\be
\BG' = (\BA_i\Bx_i+\Bx_i\BA_i)
+\alpha\BA_i\BA_i=0\ .
\ee
Then in this construction, the gauge field has 
correct degrees of freedom, even in the
noncommutative case.
Apparently, this theory also gives the correct commutative limit.

To complete our discussion, we want to mention that the use of the 
constraint $\BG'=0$ to restrict the differential calculus 
is not straightforward, since 
$d\BG'$ does not automatically vanish. 
The treatment of this constraint within the 
differential calculus needs more investigation.

The fuzzy sphere is one of the easiest examples of a 
noncommutative space. We can consider the $U(1)$ 
gauge theory on the 
fuzzy sphere formulated in this paper as a regularized version of a 
gauge theory on the sphere.  The gauge theory on the noncommutative 
sphere is also investigated in ref.\cite{Klimcik97}.  
The differential calculus
there is based on the supersymmetric fuzzy sphere 
and the structure of the
fermion is different from the one discussed here. 
Thus the structure of the differential algebra is
also different. However, this is not a contradiction since,
 in principle, 
there are many types of differential algebra
associated with the fuzzy sphere algebra, depending on the
choice of the spectral triple.

In the formulation given here
we can also see an interesting analogue with the M(atrix) theory.
If we introduce a new field
\be
\nabla_i={1\over\alpha}\Bx_i+\BA_i\ ,
\ee
then the field strength $\Theta_{ij}^A$ is given by
\be
\Theta_{ij}^A=i[\nabla_i,\nabla_j]
-\epsilon_{ijk}\nabla_k\ .
\ee
Using the same replacement for the symmetric part, the action is
\be
\BS_G={1\over(N+1)}\tr_{\CF}\{C_1\Big(i[\nabla_i,\nabla_j]
-\epsilon_{ijk}\nabla_k\Big)^2+C_2(\nabla_i\nabla_i-{\ell^2\over \alpha^2})\}\ .\label{MatrixAction}
\ee

After rewriting the gauge field action in the above form, 
we can make the following reinterpretation:
There is a general theory defined by the matrix $\nabla_i$ 
and the action (\ref{MatrixAction}).  
The geometry of the base space is then
defined by the vacuum expectation value of the field 
$\nabla_i$
 given by $\langle\nabla_i\rangle={\Bx_i\over\alpha}$.
Then the original gauge field action can be obtained by
 expanding the field around this vacuum expectation value.

\bigskip
\noindent{\Large \bf{Acknowledgement}}
\medskip

The authors would like to thank H. Ishikawa for helpful discussions. 
This work is supported by the Grant-in-Aid of Monbusho (the Japanese Ministry of Education, Science, Sports and Culture) \#09640331.

\section{Appendix}

\subsection{One form kernels}

We show the existence of nontrivial elements of $\BJ^{(1)}$ which 
contribute to $\pi(d ker\pi^{(1)})$. 
Consider the $1$-forms:
\be
(\Bx_A)^pd(\Bx_B)^q  \label{kernel_omega}
\ee
with $A,B=+,-,3$, 
where we have used the coordinates $\Bx_\pm=\Bx_1\pm i\Bx_2$. 
Since $\CA_N$ is the algebra of $(N+1)\times(N+1)$ matrices,
corresponding to the $(N+1)$ dimensional representation 
of the algebra of the angular momentum up to 
the normalization, the identity 
$(\Bx_\pm)^{N+1}=0$ holds, and thus one 
easily finds that elements of the 
differential kernel appear for $A=B=\pm$. 
Here, we give the proof for $A=B=+$. (The proof for 
$A=B=-$ works correspondingly.)

Let us define the $1$-forms $\omega_{p,q}$ as
\be
\omega_{p,q}=\Bx_+^pd\Bx_+^q\ .
\ee
then the following proposition holds.

\proposition{5}
$\omega_{p,q}$ is an element of $ker\pi^{(1)}$, 
for integers $p,q$ satisfying $1<p,q<N+1$ and 
 $p+q\geq N+2$.

\proof
Using the Dirac operator given in (\ref{Dirac0}) we obtain
\be
\pi(\omega_{p,q})=\Bx_+^p[\BD,\Bx_+^q]={i\over \ell}
\sum_{A=+,3,-}\gamma_\chi\chi^o_A\Bx_+^p\BL_A\Bx_+^q \label{kernel1}
\ee
A straightforward calculation yields 
\bea
L_+\Bx_+^q&=&0\ ,\cr
L_3\Bx_+^q &=&q\Bx_+^{q}\ ,\cr
L_-\Bx_+^q &=&-2q\Bx_+^{q-1}(\Bx_3+{\alpha\over 2}(q-1))\ .
\label{DerivativeOfxPlus}
\eea
Substituting the above relations, and using that $\Bx_+^{N+1}=0$,
the r.h.s. of eq.(\ref{kernel1}) vanishes.
\qed

Note that there are six different elements 
$\Bx_\lambda$ which correspond to the 
raising (lowering) operators of the three different
directions $\Bx_\lambda=\Bx_j\pm i\Bx_k$, where $j<k$ and 
$j,k\in\{1,2,3\}$, satisfying
\be
(\Bx_\lambda)^{N+1}=0.           \label{id}
\ee
For each direction $\Bx_\lambda$ 
we can obtain kernels of the type 
$\omega_{p,q}=\Bx_\lambda^pd\Bx_\lambda^q$.\footnote{
In fact we have a whole 'tower' of kernels 
\be
\prod_{k=0}^m(\Bx_3-{(N-2k)\over 2}\alpha)\omega_{p,q}\in ker\pi^{(1)} \ , \ 
{\rm for}\ p+q\geq N+1-m \ ,\ m=0,...,N-1 \ ,
\ee
since 
$\prod_{k=0}^m(\Bx_3-{(N-2k)\over 2}\alpha)\Bx_+^{N-m}=0$. 
However the above kernel 
$\omega_{p,q}$ is enough for the following discussions.}
 These one forms 
as well as all one forms obtained by 
multiplying elements $a\in\CA_N$ onto 
them, 
belong to the kernel $\BJ^{(1)}=ker\pi^{(1)}$.

We may still find other elements of $ker\pi^{(1)}$. However, the
above kernel $\omega_{p,q}$ is
sufficient to 
prove that $\pi(d\, ker\pi^{(1)})$ is not empty and contains 
the symmetric traceless part of $\chi^o_i\chi^o_j$.

\end{document}